%% file: lanchester_arxiv.tex
\documentclass{article}

\usepackage{arxiv}

\usepackage[utf8]{inputenc} 
\usepackage[T1]{fontenc}    
\usepackage{hyperref}       
\usepackage{url}            
\usepackage{booktabs}       
\usepackage{amsfonts}       
\usepackage{amssymb}        
\usepackage{amsmath}        
\usepackage{nicefrac}       
\usepackage{microtype}      
\usepackage{lipsum}		
\usepackage{graphicx}
\usepackage[round]{natbib}
\usepackage{doi}
\usepackage{wrapfig}

\title{Discrete Lanchester attrition models: the case of precautionary surrender}

\author{ \href{https://orcid.org/0000-0001-5982-0415}{\includegraphics[width=0.03\textwidth]{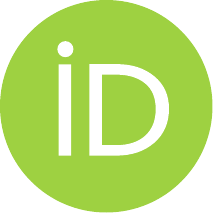}\hspace{1mm}Robin K. S.~Hankin}\thanks{\href{https://academics.aut.ac.nz/robin.hankin}{work};  
\href{https://www.youtube.com/watch?v=JzCX3FqDIOc&list=PL9_n3Tqzq9iWtgD8POJFdnVUCZ_zw6OiB&ab_channel=TrinTragulaGeneralRelativity}{play}} \\
  University of Stirling\\
  \texttt{hankin.robin@gmail.com}\\
}

\hypersetup{
pdftitle={Naval warfare in R},
pdfsubject={q-bio.NC, q-bio.QM},
pdfauthor={Robin K. S.~Hankin},
pdfkeywords={Naval warfare, stochastic model}
}

\usepackage{tikz}
\usepackage{Sweave}
\begin{document}
\maketitle

\begin{abstract}

  Discrete Lanchester-type attrition models describe many types of
  antagonistic situations; the preferred interpretation is two fleets
  of battleships, each trying to sink the other.  Such models may be
  characterised by a bivariate recurrence relation.  Here I consider a
  restricted case in which a fleet that finds itself two or three
  units behind its opponent immediately surrenders.  I present some
  theoretical and numerical results and suggest lines for further
  work.

\end{abstract}

\section{Introduction}

Motivated by the analysis of modern and ancient military strategy,
\citet{lanchester1956} set out a simple model for two mutually hostile
forces, which he styled ``red'' and ``blue'', of strength $r=r(t)$ and
$b=b(t)$ respectively:

\begin{equation}\label{lanchester}
\begin{split}
\frac{dr}{dt}  &= -bk\\
\frac{db}{dt}  &= -rc.
\end{split}
\end{equation}

Here, $c$ and $k$ are constants representing the military strength of
the individual units; $c=k$ if the fighting values of the individual
units of the force are equal.  Lanchester went on to study the
properties of this system and considered British naval tactics in the
eighteenth and nineteenth centuries in the light of his equations.
\citet{ancker1987} considered
equations~\ref{lanchester} and presented extensive numerical and
theoretical results.  It is interesting to compare this with the
system semi-discrete system considered by~\citet{fox2010}, who assumed
that the system was discrete in time but not fighting strength; in his
work, the fighting strengths of the nations were conceptualised as
functions $x(n)$ and $y(n)$ from the integers to the non-negative
reals.  His analogues to equations~\ref{lanchester} were

\begin{equation}
\begin{split}
x(n+1) &= x(n) -k_xy(n)\\
y(n+1) &= y(n) -k_yx(n),\qquad n=0,1,2,\ldots.
\end{split}
\end{equation}

Writing in 1974, \citeauthor{karr1974} considered a number of similar
Lanchester-type situations, focussing on the continuous case, and
presented several ODEs.  He went on to consider (p41) the case where
the fighting units are discrete, being destroyed according to a
stochastic process.  His model may be expressed as a recursive scheme
for $p(r,b)$, $r,b\in\mathbb{N}\backslash (0,0)$:

\begin{equation}\label{defining}
p(r,b) = \begin{cases}
1 & r\geqslant 1, b=0\\
0 & b\geqslant 1, r=0\\
\frac{r}{r+b}\cdot p(r,b-1) + \frac{b}{r+b}\cdot p(r-1,b)  & r,b\geqslant 1.
\end{cases}
\end{equation}

This describes the OK Corral
shootout~\citep{williams1998,kingman2003}, although here the preferred
interpretation of the problem is as follows: navies $R,B$ deploy
fleets of $r,b$ ships respectively and engage in battle.  At any point
in time, the probability that the next ship to be sunk is one of $R$
is $b/(r+b)$.  The winner is the first to destroy all the other
fleet's ships; if $(r,b)$ is the initial state vector, $p(r,b)$
becomes the probability of navy $R$ winning, that is, achieving state
$(b,0)$ for some $b>0$.

From now on, it is assumed that $r,b,n\in\mathbb{N}^+$ unless
otherwise indicated.  The following results are immediate:

\begin{itemize}
\item $0<p(r,b)<1$, $r,b>0$
\item $p(r,b)+p(b,r)=1$
\item $p(n,n)=\frac{1}{2}$, $n>0$
\item $p(n,1)=\frac{(n+1)!-1}{(n+1)!}$, $p(1,n)=\frac{1}{(n+1)!}$, $n>0$.
\end{itemize}

\section{Results}

\begin{figure}[p]
\begin{center}
\includegraphics{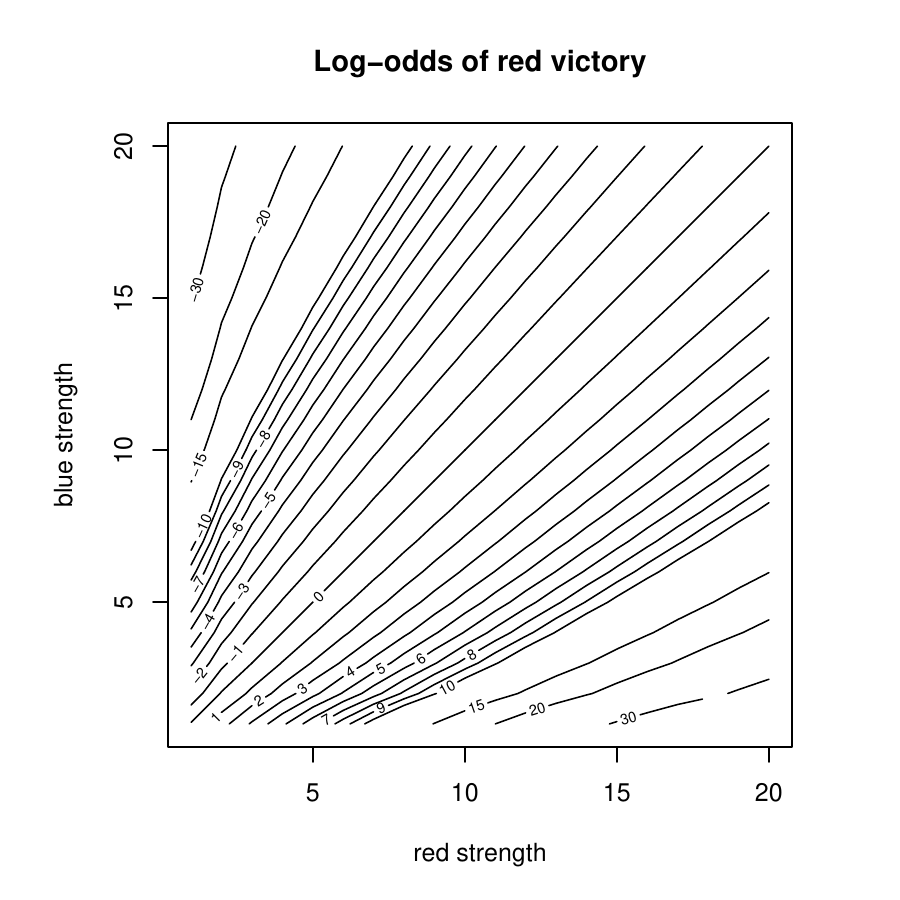}
\caption{log-odds \label{logoddsratioredvsblue} ratio ${\log}{\left(\frac{p}{1-p}\right)}$ for red victory as a
function of red and blue strength $1(1)20$}
\end{center}
\end{figure}

Figure~\ref{small_ab} shows a graphical representation of
equations~\ref{defining} together with win probabilities for small
values of $(r,b)$; and figures~\ref{probrounda}
and~\ref{logoddsratioredvsblue} show win probabilities for a wider
range of initial states.

\usetikzlibrary{arrows.meta}
\usetikzlibrary{patterns}
\newcommand{\arrowIn}{
\tikz \draw[-{Latex[length=4mm]}] (-1pt,0) -- (1pt,0);
}

\begin{figure}[p]
\begin{center}
\input{small_ab_transition.tex}\caption{Transition diagram for $(r,b)\preccurlyeq(4,3)$.
Red\label{small_ab} figures refer to the red forces and blue to the
blue forces.  Round brackets indicate the state of each node; thus
$(\textcolor{red}{3},\textcolor{blue}{4})$ means 3 red and 4 blue
extant fighting units.  Angle brackets indicate the probabilities for
red and blue victory; thus
$\left\langle\textcolor{red}{\frac{233}{315}},\textcolor{blue}{\frac{82}{315}}\right\rangle$
means that the probability of (eventual) red victory is
$\frac{233}{315}$ and that of blue victory $\frac{82}{315}$.  Teal
numbers indicate transition probabilities
$\frac{r}{r+b},\frac{b}{r+b}$; large dots show absorbing states}
\end{center}
\end{figure}

\begin{figure}[p]
\begin{center}
\begin{Schunk}
\begin{Soutput}
   blue
red     1     2     3     4     5     6     7     8     9
  1 0.500 0.167 0.042 0.008 0.001 0.000 0.000 0.000 0.000
  2 0.833 0.500 0.225 0.081 0.024 0.006 0.001 0.000 0.000
  3 0.958 0.775 0.500 0.260 0.113 0.042 0.013 0.004 0.001
  4 0.992 0.919 0.740 0.500 0.285 0.139 0.059 0.022 0.008
  5 0.999 0.976 0.887 0.715 0.500 0.303 0.161 0.076 0.032
  6 1.000 0.994 0.958 0.861 0.697 0.500 0.317 0.179 0.091
  7 1.000 0.999 0.987 0.941 0.839 0.683 0.500 0.329 0.195
  8 1.000 1.000 0.996 0.978 0.924 0.821 0.671 0.500 0.338
  9 1.000 1.000 0.999 0.992 0.968 0.909 0.805 0.662 0.500
\end{Soutput}
\end{Schunk}
\caption{Probabilities of red \label{probrounda} victory from initial states $r,b=1(1)9$}
\end{center}
\end{figure}

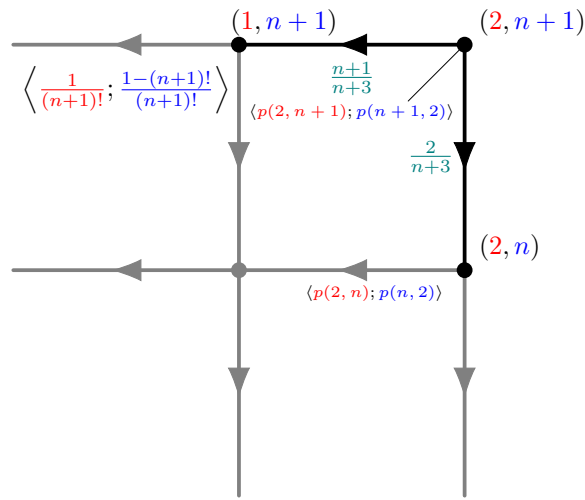
\begin{figure}[p]
\begin{center}
\input{single_cell.tex}\caption{Recurrence relation \label{single_cell2} cell analysis, colour coding as
for figure~\ref{small_ab}}
\end{center}
\end{figure}

Figure~\ref{single_cell2} shows the general case for $r=2$.  This has
recurrence relation $x_{n+1}=x_n\frac{n+1}{n+3} +
\frac{1}{(n+2)!}\cdot\frac{2}{n+3}$ which has solution

\begin{equation}\label{xn}
p(2,n) = x_n =
\frac{2^{n+2}-(n+2)}{(n+2)!},\qquad n\geqslant 0
\end{equation}

leading to
\begin{equation}\label{xnp2}
p(2,t) = 
\frac{2^{t}-t}{t!}\sim
\frac{1}{\sqrt{2\pi t}}\cdot\left(\frac{2e}{t}\right)^t,\qquad t\geqslant 2
\end{equation}

where $t$ is the total number of ships; proof by direct substitution.
The asymptotic form is via Stirling's
approximation~\citep[p257]{abramowitz1965}.  We may verify
equation~\ref{xn} by cross-checking it against figure~\ref{small_ab},
using the R computer language~\citep{rcore2023}:

\begin{Schunk}
\begin{Sinput}
> f <- function(n){(2^(n+2)-(n+2))/factorial(n+2)}
> f(1:4) - c(5/6,1/2,9/40,29/360)
\end{Sinput}
\begin{Soutput}
[1] 0 0 0 0
\end{Soutput}
\end{Schunk}

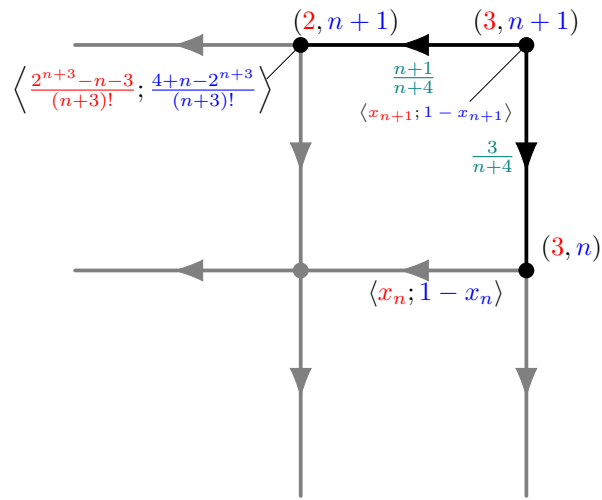
\begin{figure}[p]
\begin{center}
\input{single_cell3.tex}\caption{Single\label{single_cell3} cell analysis}
\end{center}
\end{figure}

Above we see agreement between equation~\ref{xn} and
Figure~\ref{small_ab}.  Proceeding to consider $r=3$
(Figure~\ref{single_cell3}), we obtain the relation
$x_{n+1}=x_n\cdot\frac{3}{n+4} +
\frac{2^{n+3}-n-3}{(n+3)!}\cdot\frac{n+1}{n+4}$ which, together with
$x_0=1$, gives

\begin{equation}\label{xn3}
p(3,n)=x_n=\frac{54\cdot
3^n-16(n+3)\cdot 2^n+(n+3)(n+2)}{2(n+3)!}
\sim\frac{27}{\sqrt{2\pi n^7}}\left(\frac{3e}{n}\right)^n.
\end{equation}

The general pattern does not seem at all easy to obtain.  However, it
might be reasonable to conjecture that
$p(n,b)\sim\alpha\cdot(be/n)^n\cdot n^{-(b+1/2)}$ for some $\alpha$ as
$n\longrightarrow\infty$, with $b$ fixed.

\subsection{The restricted game: precautionary surrender}

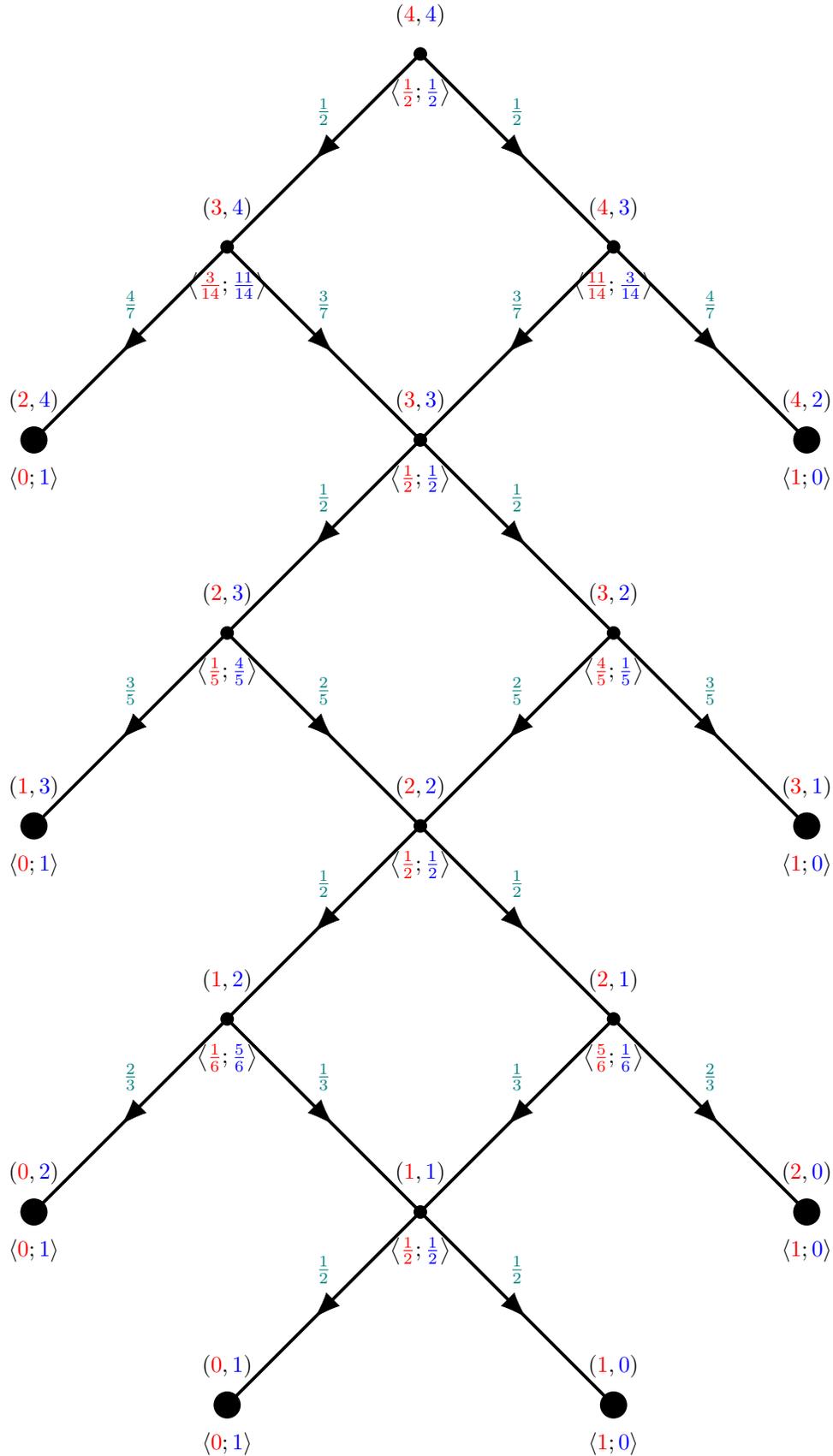
\begin{figure}
\begin{center}
\input{nnp1.tex}\caption{Win probabilities for the restricted game: \label{nnp1} being
two or more units behind means surrender}
\end{center}
\end{figure}

\begin{figure}[p]
\begin{center}
\input{nnp1_rec.tex}\caption{Single\label{nnp1_rec} cell analysis for the restricted game, $n>1$}
\end{center}
\end{figure}
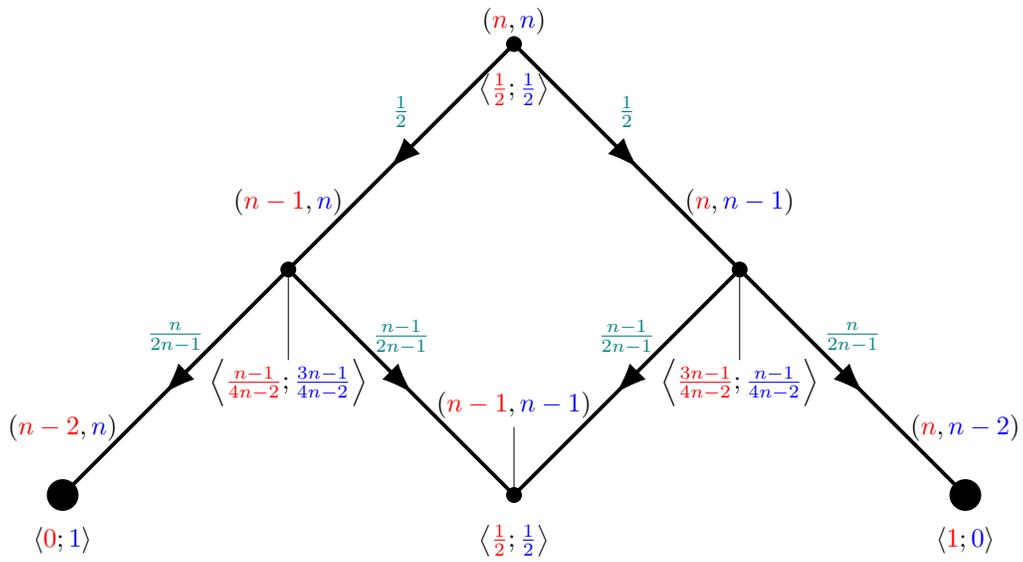

Given that $p(n,n)=1/2$, one might seek solutions for $p(n,n+1)$ but
this seems to be not at all easy.  However, we may consider a simpler
game in which we specify $p(n,n+2)=0$ and $p(n+2,n)=1$ for all
$n\geqslant 0$: the game is considered decided if
$\left|r-b\right|\geqslant 2$ at any time.  One interpretation might
be that a navy surrenders if it finds itself two or more units behind
its opponent; figure~\ref{nnp1} shows some results for small values of
$(r,b)$.  The situation is not quite isomorphic to that of tennis at
deuce because the restricted game has a pre-specified length, while a
tennis game has no upper number of points.  From figure~\ref{nnp1_rec}
we conclude that

\begin{equation}
p(n-1,n)=\begin{cases}
\frac{3n-1}{4n-2} &        n>1  \\
0                 & \mbox{$n=1$.}
\end{cases}
\end{equation}

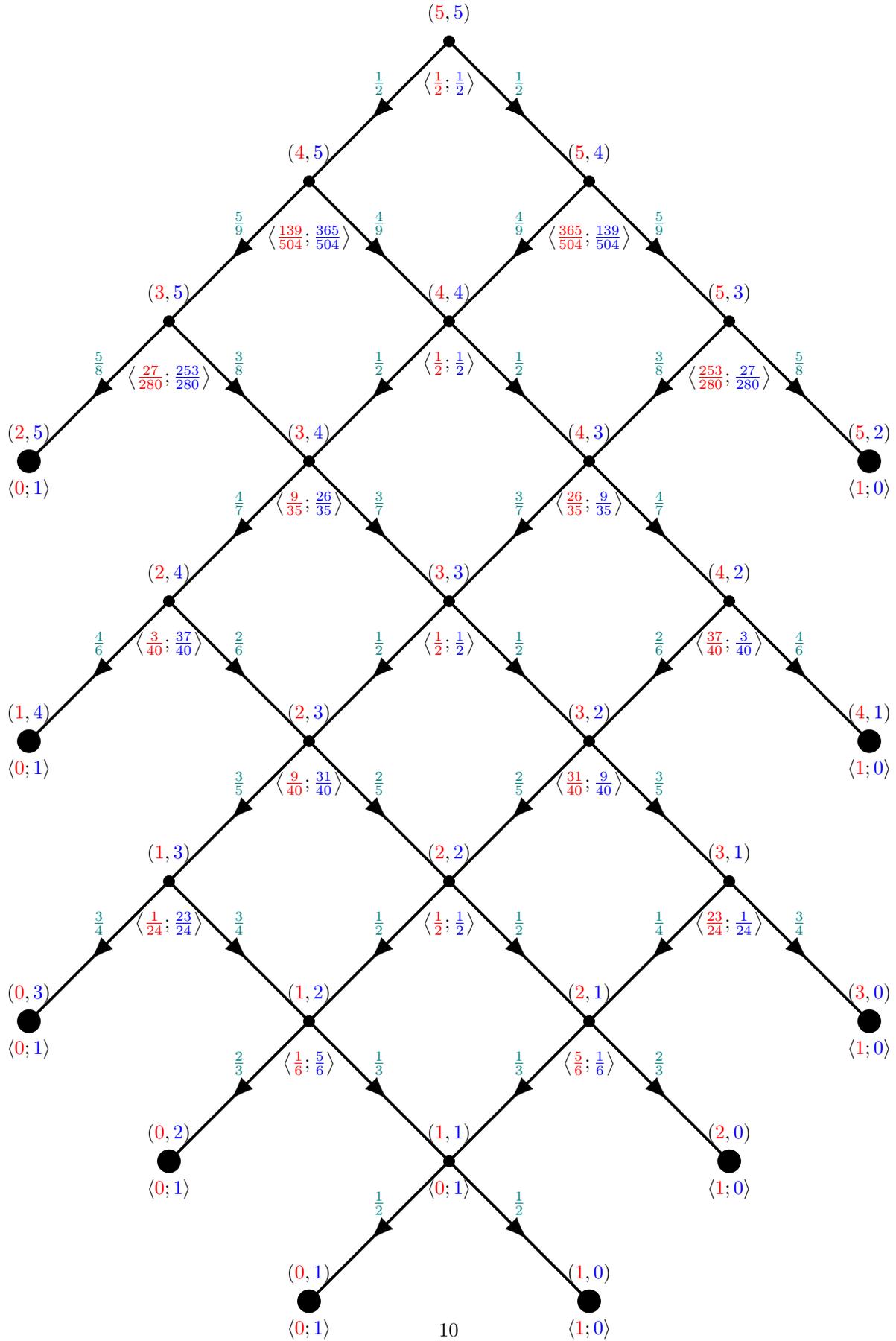
\begin{figure}[p]
\begin{center}
\input{nnp2.tex}\caption{Restricted game: \label{nnp2} being {\em three} or more
behind means surrender}
\end{center}
\end{figure}

\begin{figure}[p]
\begin{center}
\input{nnp2_cell.tex}\caption{Precautionary surrender on being three or more units
behind,\label{nnp2_cell} $n>1$}
\end{center}
\end{figure}
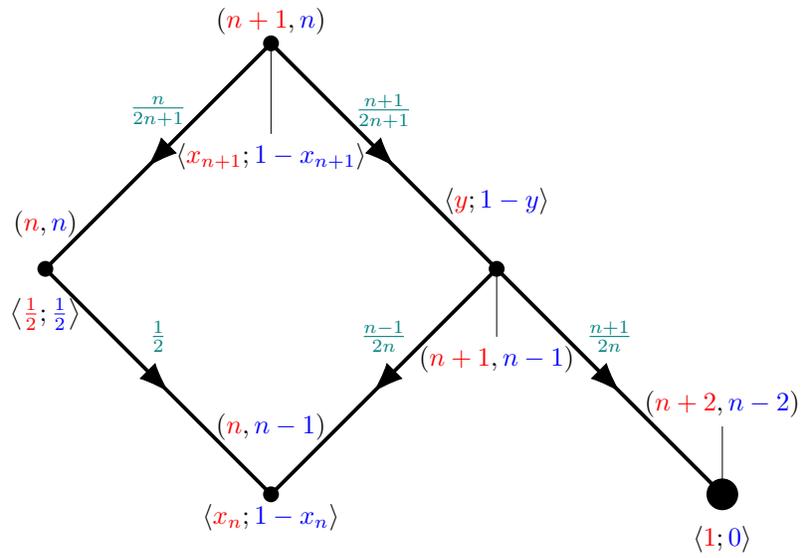

The next natural game to consider is that a navy surrenders if it
finds itself three or more units down; figure~\ref{nnp2}.
Figure~\ref{nnp2_cell} shows a single cell and gives us the relation
$x_{n+1} = \frac{1}{2}\cdot\frac{n}{2n+1} + y\cdot\frac{n+1}{2n+1}$
with $y = x_n\cdot\frac{n-1}{2n} + 1\cdot\frac{n+1}{2n}$, or
altogether

\begin{equation}\label{recurrence_prec}
x_{n+1}= x_n\cdot\frac{n^2-1}{4n^2+2n} + \frac{2n^2+2n+1}{4n^2+2n}.
\end{equation}

Although involved and complicated, this equation is solved easily by
computer algebra systems.  Sage gives us

\begin{equation}\label{sage}
\frac{
	3\cdot 4^{n}n(n-2)!\sum_{j=2}^{n-1}\frac{
	2(2j^2+2j+1)\cdot 4^{j-n}\left(j+\frac{1}{2}\right)!
	}{
	(2j+1)(j+1)!
	}
	+ 5\sqrt{\pi}n(n-2)!}{
4^n\cdot\left(n-\frac{1}{2}\right)!
}
\end{equation}

and it is straightforward but tedious to verify that
equation~\ref{sage} satisfies recurrence
relation~\ref{recurrence_prec}.  Mathematica gives us a more opaque
but more easily evaluated equivalent form

\newcommand{\Fmn}[2]{\operatorname{{}_{#1}F_{#2}}}
\newcommand{\ft}{\Fmn{2}{1}}
\newcommand{\four}[4]{\frac{\Gamma\left(#1\right)\Gamma\left(#2\right)}{\Gamma\left(#3\right)\Gamma\left(#4\right)}}

\begin{equation}\label{mathematica}
\frac{1}{18(n-1)}
\left(
12n-8- \frac{(9+i\sqrt{3})\sqrt{\pi}\Gamma(n+1)}{4^n\Gamma(n+1/2)}-2\Gamma(n+1)\widetilde{{}_2F_1}(1,n+1/2;n+2;4)
\right)
\end{equation}

(the tilde in $\widetilde{{}_2F_1}(1,n+1/2;n+2;4)$ means the {\em
regularized} ${}_2F_1$ Gauss hypergeometric
function~\citep{abramowitz1965}).  The hypergeometric term is
problematic.  First, note that it is evaluated {\em on} the cut line
(which goes from $1+0i$ to $\infty$ along the real axis); the
hypergeometric function is defined on the cut line as being continuous
from below.  In any event, the imaginary component tends to zero as
$n\longrightarrow\infty$.  Equation 15.3.7 of~\citep{abramowitz1965},
viz

\begin{eqnarray*}\label{ams15.3.7}
\ft\left(a,b;c;z\right) = \four{c}{b-a}{b}{c-a}\left(-z\right)^{-a}\ft\left(a,1-c+a;1-b+a;\frac{1}{z}\right)
\nonumber\\
+\four{c}{a-b}{a}{c-b}\left(-z\right)^{-b}\ft\left(b,1-c+b;1-a+b;\frac{1}{z}\right)
\end{eqnarray*}

may be used to ensure that $\ft()$ is only evaluated inside the unit
circle, thereby avoiding the need to consider cut lines.  This leads
to, after simplification:

\begin{equation}\label{simplifiedversion}
p(n,n-1)=
\frac{1}{9(n-1)}
\left(6n-4-  \frac{9}{2{2n\choose n\,\,n}}+\frac{\ft(1,-n;3/2-n;1/4)}{2(2n-1)}
\right),\qquad n>1.
\end{equation}

We may verify that equation~\ref{simplifiedversion} is correct
numerically, using the R computer language~\citep{rcore2023} and the
{\tt hypergeo} package~\citep{hankin2015}:

\begin{Schunk}
\begin{Sinput}
> library("hypergeo")
> f <- function(n){
+ (6*n-4-9/(2*choose(2*n,n)) + Re(hypergeo(1,-n,3/2-n,1/4)/(2*(2*n-1)))) / (9*(n-1))
+ }
> f(2:4) - c(5/6, 31/40, 26/35)
\end{Sinput}
\begin{Soutput}
[1] 0 0 0
\end{Soutput}
\end{Schunk}

\begin{figure}[p]
  \begin{center}
\begin{Schunk}
\begin{Sinput}
> library("hypergeo")
> f1 <- function(n){Re(hypergeo(1,-n,3/2-n,1/4))}
> f2 <- function(n){4*(1+0.5/n+1/n^2)/3 }
> n <- 10:500
> plot(n,abs(f1(n)-f2(n)),log="y")
\end{Sinput}
\end{Schunk}
\includegraphics{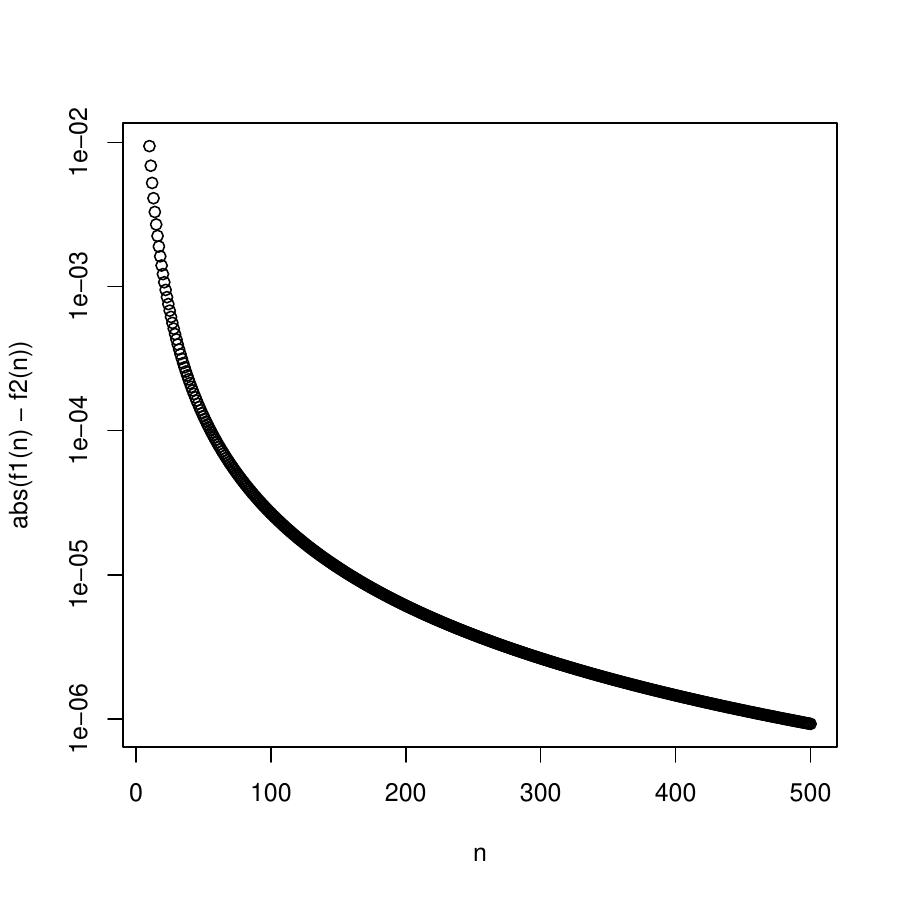}
\caption{Difference\label{hypergeoasymp} between $\ft(1,-n;3/2-n;1/4)$
and $\frac{4}{3} +\frac{1}{3}n^{-1}+\frac{4}{3}n^{-2}$ for
$10\leqslant n\leqslant 500$, logarithmic scale.  We see that the
exact and approximate results for $n=500$ differ by about
$10^{-6}$}
\end{center}
\end{figure}

Above we see agreement between equation~\ref{simplifiedversion} and
figure~\ref{small_ab}.  Even in its streamlined form, the
hypergeometric term remains troublesome.  Figure~\ref{hypergeoasymp}
presents numerical evidence to suggest that

\begin{equation}\label{noproof}
\ft(1,-n;3/2-n;1/4)=\frac{4}{3} +
\frac{2}{3}n^{-1} + \frac{4}{3}n^{-2} + \mathcal{O}(n^{-3}).
\end{equation}

\begin{figure}[p]
\begin{center}
\input{nnp3_cell.tex}\caption{Precautionary surrender on being four or more units
behind,\label{nnp3_cell} $n>1$}
\end{center}
\end{figure}
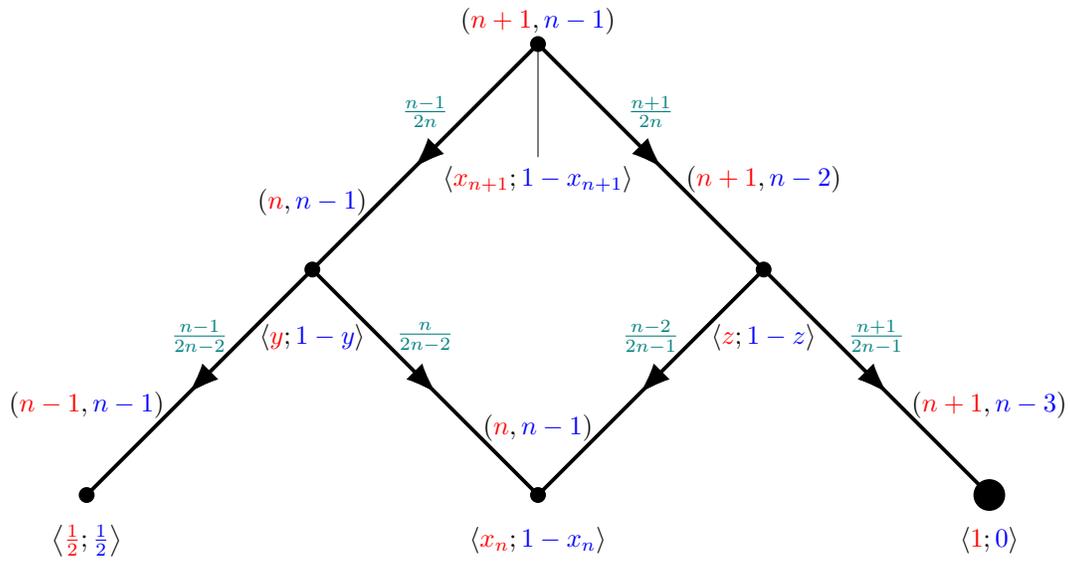

However, an algebraic proof that
$\displaystyle\lim_{n\longrightarrow\infty}\ft(1,-n;3/2-n;1/4)$ even
exists is not available; but equation~\ref{noproof} would imply that
$p(n,n-1)=\frac{2}{3} + \frac{2}{9}n^{-1} +\frac{7}{27}n^{-2}
+\mathcal{O}{\left(n^{-3}\right)}$.  The case for precautionary
surrender on being four or more units behind is shown in
figure~\ref{nnp3_cell}.  The recurrence would be

\begin{equation}
\begin{split}
p(n,n-1) =\label{fourbehind}
x_n &= \frac{n-1}{2n}\cdot y + \frac{2n+1}{2n}\cdot z\\
y   &= \frac{n-1}{2n-1}\cdot\left(\frac{1}{2}\right) + \frac{n}{2n-1}\cdot x_n\\
z   &= \frac{n+1}{2n-1}\cdot x_n +\frac{n-2}{2n-1}\cdot 1.
\end{split}
\end{equation}

However, equation~\ref{fourbehind} does not appear to have a useful analytic
solution.

\section{Conclusions and further work}

Discrete Lanchester systems appear to be an interesting problem but
exact solutions are elusive.  Here winning probabilities for the full
problem were presented for initial state vectors with small entries;
the general case for $(2,n)$ and $(3,n)$ was given.  Asymptotic
results for a restricted problem (precautionary surrender) were given
where numerical superiority of two or three units resulted in
pre-emptive surrender of the behind side.

Military interpretations of discrete Lanchester models would suggest
that, for given $n$ and $\alpha$, finding the minimum value of $a$
such that $p(n,n+a)\geqslant\alpha$ would have direct significance.
One would then be seeking the numerical superiority required to secure
victory with a given probability.  In general, one might hope to
discover discrete analogues of Lanchester's N-square
law~\citep{lanchester1956,lepingwell1987}.
Figure~\ref{logoddsratioredvsblue} goes some way to answering this,
but asymptotic results would be interesting.

Such questions give rise to a natural generalization of discrete
Colonel Blotto-type zero-sum games~\citep{hart2008}: how might one
allocate resources among battlefields to maximize the expected number
of victories?  Further, one might regard one's combat units as
vulnerable targets rather than offensive assets; and consider their
placement from the perspective of liability allocation
games~\citep{hankin2020}.

One natural generalization, of interest all the way back to
Lanchester's seminal 1956 work, is to consider the case where the
ships of each fleet have different effectivenesses $\alpha,\beta$,
leading to probabilities $\frac{\alpha r}{\alpha r+\beta b},
\frac{\beta b}{\alpha r+\beta b}$.  Results for this general case seem
to be not at all easy to obtain, beyond producing diagrams analogous
to figures~\ref{small_ab} and~\ref{logoddsratioredvsblue}.
Restricting attention to equal fighting strengths, one might observe
that not only is the eventual winner a random variable, but so too is
the strength of the remaining fleet at victory.  Investigating the
probability mass function for this might be interesting.

\clearpage
\bibliographystyle{plainnat}
\bibliography{naval}

\end{document}

%% file: small_ab_transition.tex
\begin{tikzpicture}[line cap=round,line join=round,x=4cm,y=4cm]

\newcommand{\state}[2]{$(\textcolor{red}{#1},\textcolor{blue}{#2})$}
\newcommand{\probs}[2]{$\left\langle\textcolor{red}{#1};\textcolor{blue}{#2}\right\rangle$}

\fill (1,0) circle[radius=6pt];
\fill (2,0) circle[radius=6pt];
\fill (3,0) circle[radius=6pt];
\fill (4,0) circle[radius=6pt];
\fill (0,1) circle[radius=6pt];
\fill (1,1) circle[radius=3pt];
\fill (2,1) circle[radius=3pt];
\fill (3,1) circle[radius=3pt];
\fill (4,1) circle[radius=3pt];
\fill (0,2) circle[radius=6pt];
\fill (1,2) circle[radius=3pt];
\fill (2,2) circle[radius=3pt];
\fill (3,2) circle[radius=3pt];
\fill (4,2) circle[radius=3pt];
\fill (0,3) circle[radius=6pt];
\fill (1,3) circle[radius=3pt];
\fill (2,3) circle[radius=3pt];
\fill (3,3) circle[radius=3pt];
\fill (4,3) circle[radius=3pt];

\draw[line width=0.5mm] (1,1) -- (0,1) node[sloped, pos=0.5, allow upside down]{\arrowIn};
\draw[line width=0.5mm] (2,1) -- (1,1) node[sloped, pos=0.5, allow upside down]{\arrowIn};
\draw[line width=0.5mm] (3,1) -- (2,1) node[sloped, pos=0.5, allow upside down]{\arrowIn};
\draw[line width=0.5mm] (4,1) -- (3,1) node[sloped, pos=0.5, allow upside down]{\arrowIn};
\draw[line width=0.5mm] (1,2) -- (0,2) node[sloped, pos=0.5, allow upside down]{\arrowIn};
\draw[line width=0.5mm] (2,2) -- (1,2) node[sloped, pos=0.5, allow upside down]{\arrowIn};
\draw[line width=0.5mm] (3,2) -- (2,2) node[sloped, pos=0.5, allow upside down]{\arrowIn};
\draw[line width=0.5mm] (4,2) -- (3,2) node[sloped, pos=0.5, allow upside down]{\arrowIn};
\draw[line width=0.5mm] (1,3) -- (0,3) node[sloped, pos=0.5, allow upside down]{\arrowIn};
\draw[line width=0.5mm] (2,3) -- (1,3) node[sloped, pos=0.5, allow upside down]{\arrowIn};
\draw[line width=0.5mm] (3,3) -- (2,3) node[sloped, pos=0.5, allow upside down]{\arrowIn};
\draw[line width=0.5mm] (4,3) -- (3,3) node[sloped, pos=0.5, allow upside down]{\arrowIn};

\draw[line width=0.5mm] (1,1) -- (1,0) node[sloped, pos=0.5, allow upside down]{\arrowIn};
\draw[line width=0.5mm] (1,2) -- (1,1) node[sloped, pos=0.5, allow upside down]{\arrowIn};
\draw[line width=0.5mm] (1,3) -- (1,2) node[sloped, pos=0.5, allow upside down]{\arrowIn};
\draw[line width=0.5mm] (2,1) -- (2,0) node[sloped, pos=0.5, allow upside down]{\arrowIn};
\draw[line width=0.5mm] (2,2) -- (2,1) node[sloped, pos=0.5, allow upside down]{\arrowIn};
\draw[line width=0.5mm] (2,3) -- (2,2) node[sloped, pos=0.5, allow upside down]{\arrowIn};
\draw[line width=0.5mm] (3,1) -- (3,0) node[sloped, pos=0.5, allow upside down]{\arrowIn};
\draw[line width=0.5mm] (3,2) -- (3,1) node[sloped, pos=0.5, allow upside down]{\arrowIn};
\draw[line width=0.5mm] (3,3) -- (3,2) node[sloped, pos=0.5, allow upside down]{\arrowIn};
\draw[line width=0.5mm] (4,1) -- (4,0) node[sloped, pos=0.5, allow upside down]{\arrowIn};
\draw[line width=0.5mm] (4,2) -- (4,1) node[sloped, pos=0.5, allow upside down]{\arrowIn};
\draw[line width=0.5mm] (4,3) -- (4,2) node[sloped, pos=0.5, allow upside down]{\arrowIn};

\node at (1.2, 0.1)  {\state{1}{0}};
\node at (2.2, 0.1)  {\state{2}{0}};
\node at (3.2, 0.1)  {\state{3}{0}};
\node at (4.2, 0.1)  {\state{4}{0}};
\node at (0.2, 1.1)  {\state{0}{1}};
\node at (1.2, 1.1)  {\state{1}{1}};
\node at (2.2, 1.1)  {\state{2}{1}};
\node at (3.2, 1.1)  {\state{3}{1}};
\node at (4.2, 1.1)  {\state{4}{1}};
\node at (0.2, 2.1)  {\state{0}{2}};
\node at (1.2, 2.1)  {\state{1}{2}};
\node at (2.2, 2.1)  {\state{2}{2}};
\node at (3.2, 2.1)  {\state{3}{2}};
\node at (4.2, 2.1)  {\state{4}{2}};
\node at (0.2, 3.1)  {\state{0}{3}};
\node at (1.2, 3.1)  {\state{1}{3}};
\node at (2.2, 3.1)  {\state{2}{3}};
\node at (3.2, 3.1)  {\state{3}{3}};
\node at (4.2, 3.1)  {\state{4}{3}};

\node[text=teal] at (0.5, 0.85)  {$\frac{1}{2}$};
\node[text=teal] at (1.5, 0.85)  {$\frac{1}{3}$};
\node[text=teal] at (2.5, 0.85)  {$\frac{1}{4}$};
\node[text=teal] at (3.5, 0.85)  {$\frac{1}{5}$};
\node[text=teal] at (0.5, 1.85)  {$\frac{2}{3}$};
\node[text=teal] at (1.5, 1.85)  {$\frac{2}{4}$};
\node[text=teal] at (2.5, 1.85)  {$\frac{2}{5}$};
\node[text=teal] at (3.5, 1.85)  {$\frac{2}{6}$};
\node[text=teal] at (0.5, 2.85)  {$\frac{3}{4}$};
\node[text=teal] at (1.5, 2.85)  {$\frac{3}{5}$};
\node[text=teal] at (2.5, 2.85)  {$\frac{3}{6}$};
\node[text=teal] at (3.5, 2.85)  {$\frac{3}{7}$};

\node[text=teal] at (0.85, 0.5)  {$\frac{1}{2}$};
\node[text=teal] at (1.85, 0.5)  {$\frac{2}{3}$};
\node[text=teal] at (2.85, 0.5)  {$\frac{3}{4}$};
\node[text=teal] at (3.85, 0.5)  {$\frac{4}{5}$};
\node[text=teal] at (0.85, 1.5)  {$\frac{1}{3}$};
\node[text=teal] at (1.85, 1.5)  {$\frac{2}{4}$};
\node[text=teal] at (2.85, 1.5)  {$\frac{3}{5}$};
\node[text=teal] at (3.85, 1.5)  {$\frac{4}{6}$};
\node[text=teal] at (0.85, 2.5)  {$\frac{1}{4}$};
\node[text=teal] at (1.85, 2.5)  {$\frac{2}{5}$};
\node[text=teal] at (2.85, 2.5)  {$\frac{3}{6}$};
\node[text=teal] at (3.85, 2.5)  {$\frac{4}{7}$};

\node at ( 0.85, -0.1){\probs{1}{0}};
\node at ( 1.85, -0.1){\probs{1}{0}};
\node at ( 2.85, -0.1){\probs{1}{0}};
\node at ( 3.85, -0.1){\probs{1}{0}};
\node at (-0.15,  0.9){\probs{0}{1}};
\node at ( 0.8 ,  0.9){\probs{\frac{1}{2}}{\frac{1}{2}}};
\node at ( 1.8 ,  0.9){\probs{\frac{5}{6}}{\frac{1}{6}}};
\node at ( 2.8 ,  0.9){\probs{\frac{23}{24}}{\frac{1}{24}}};
\node at ( 3.8 ,  0.9){\probs{\frac{119}{120}}{\frac{1}{120}}};
\node at (-0.15,  1.9){\probs{0}{1}};
\node at ( 0.8 ,  1.9){\probs{\frac{1}{6}}{\frac{5}{6}}};
\node at ( 1.8 ,  1.9){\probs{\frac{1}{2}}{\frac{1}{2}}};
\node at ( 2.8 ,  1.9){\probs{\frac{31}{40}}{\frac{9}{40}}};
\node at ( 3.8 ,  1.9){\probs{\frac{331}{360}}{\frac{29}{360}}};
\node at (-0.15,  2.9){\probs{0}{1}};
\node at ( 0.8 ,  2.9){\probs{\frac{1}{24}}{\frac{23}{24}}};
\node at ( 1.8 ,  2.9){\probs{\frac{9}{40}}{\frac{31}{40}}};
\node at ( 2.8 ,  2.9){\probs{\frac{1}{2}}{\frac{1}{2}}};
\node at ( 3.8 ,  2.9){\probs{\frac{233}{315}}{\frac{82}{315}}};
\end{tikzpicture}

%% file: single_cell.tex
\begin{tikzpicture}[line cap=round,line join=round,x=3cm,y=3cm]
\newcommand{\state}[2]{$(\textcolor{red}{#1},\textcolor{blue}{#2})$}
\newcommand{\probs}[2]{$\left\langle\textcolor{red}{#1};\textcolor{blue}{#2}\right\rangle$}

\draw[color=gray ,line width=0.5mm] (1,1) -- (0,1) node[sloped, pos=0.5, allow upside down]{\arrowIn}; 
\draw[color=gray ,line width=0.5mm] (2,1) -- (1,1) node[sloped, pos=0.5, allow upside down]{\arrowIn}; 
\draw[color=gray ,line width=0.5mm] (1,2) -- (0,2) node[sloped, pos=0.5, allow upside down]{\arrowIn}; 
\draw[color=black,line width=0.5mm] (2,2) -- (1,2) node[sloped, pos=0.5, allow upside down]{\arrowIn}; 
\draw[color=gray ,line width=0.5mm] (1,1) -- (1,0) node[sloped, pos=0.5, allow upside down]{\arrowIn}; 
\draw[color=gray ,line width=0.5mm] (1,2) -- (1,1) node[sloped, pos=0.5, allow upside down]{\arrowIn}; 
\draw[color=gray ,line width=0.5mm] (2,1) -- (2,0) node[sloped, pos=0.5, allow upside down]{\arrowIn}; 
\draw[color=black,line width=0.5mm] (2,2) -- (2,1) node[sloped, pos=0.5, allow upside down]{\arrowIn}; 

\fill[color=gray] (1,1) circle[radius=3pt];
\fill             (2,1) circle[radius=3pt];
\fill             (1,2) circle[radius=3pt];
\fill             (2,2) circle[radius=3pt];

\node at ( 0.5, 1.8)  {\probs{\frac{1}{(n+1)!}}{\frac{1-(n+1)!}{(n+1)!}}};
\node at ( 1.5, 1.7)  {{\tiny \probs{p(2,n+1)}{p(n+1,2)}}};
\node at ( 1.6, 0.9)  {{\tiny \probs{p(2,n)}{p(n,2)}}};

\node at (2.2, 1.1)  {\state{2}{n}};
\node at (1.2, 2.1)  {\state{1}{n+1}};
\node at (2.3, 2.1)  {\state{2}{n+1}};

\node[text=teal] at (1.5, 1.85)  {$\frac{n+1}{n+3}$};
\node[text=teal] at (1.85, 1.5)  {$\frac{2}{n+3}$};

\draw[line width=0.1mm] (1.75,1.75) -- (2,2);

\end{tikzpicture}

%% file: single_cell3.tex
\begin{tikzpicture}[line cap=round,line join=round,x=3cm,y=3cm]
\newcommand{\state}[2]{$(\textcolor{red}{#1},\textcolor{blue}{#2})$}
\newcommand{\probs}[2]{$\left\langle\textcolor{red}{#1};\textcolor{blue}{#2}\right\rangle$}

\draw[color=gray ,line width=0.5mm] (1,1) -- (0,1) node[sloped, pos=0.5, allow upside down]{\arrowIn};
\draw[color=gray ,line width=0.5mm] (2,1) -- (1,1) node[sloped, pos=0.5, allow upside down]{\arrowIn};
\draw[color=gray ,line width=0.5mm] (1,2) -- (0,2) node[sloped, pos=0.5, allow upside down]{\arrowIn};
\draw[color=black,line width=0.5mm] (2,2) -- (1,2) node[sloped, pos=0.5, allow upside down]{\arrowIn};
\draw[color=gray ,line width=0.5mm] (1,1) -- (1,0) node[sloped, pos=0.5, allow upside down]{\arrowIn};
\draw[color=gray ,line width=0.5mm] (1,2) -- (1,1) node[sloped, pos=0.5, allow upside down]{\arrowIn};
\draw[color=gray ,line width=0.5mm] (2,1) -- (2,0) node[sloped, pos=0.5, allow upside down]{\arrowIn};
\draw[color=black,line width=0.5mm] (2,2) -- (2,1) node[sloped, pos=0.5, allow upside down]{\arrowIn};

\draw[line width=0.1mm] (1.75,1.75) -- (2,2);
\draw[line width=0.1mm] (0.85,1.85) -- (1,2);

\fill[color=gray] (1,1) circle[radius=3pt];
\fill             (2,1) circle[radius=3pt];
\fill             (1,2) circle[radius=3pt];
\fill             (2,2) circle[radius=3pt];

\node at ( 1.6, 0.9) {\probs{x_n}{1-x_n}};
\node at ( 0.3, 1.8) {\probs{\frac{2^{n+3}-n-3}{(n+3)!}}{\frac{4+n-2^{n+3}}{(n+3)!}}};
\node at ( 1.6, 1.7) {{\tiny \probs{x_{n+1}}{1-x_{n+1}}}};

\node at (2.2, 1.1) {\state{3}{n}};
\node at (1.2, 2.1) {\state{2}{n+1}};
\node at (2.0, 2.1) {\state{3}{n+1}};

\node[text=teal] at (1.5, 1.85) {$\frac{n+1}{n+4}$};
\node[text=teal] at (1.85, 1.5) {$\frac{3}{n+4}$};

\end{tikzpicture}

%% file: nnp1.tex
\begin{tikzpicture}[line cap=round,line join=round,x=3cm,y=3cm]

\newcommand{\state}[2]{$(\textcolor{red}{#1},\textcolor{blue}{#2})$}
\newcommand{\probs}[2]{$\left\langle\textcolor{red}{#1};\textcolor{blue}{#2}\right\rangle$}

\node at (-1.0, -0.2) {\probs{0}{1}}; 
\node at ( 1.0, -0.2) {\probs{1}{0}}; 
\node at (-2   , 0.8) {\probs{0}{1}}; 
\node at (-1   , 1.8) {\probs{\frac{1}{6}}{\frac{5}{6}}}; 
\node at ( 0   , 0.8) {\probs{\frac{1}{2}}{\frac{1}{2}}}; 
\node at ( 1   , 1.8) {\probs{\frac{5}{6}}{\frac{1}{6}}}; 
\node at ( 2   , 0.8) {\probs{1}{0}}; 
\node at (-2   , 2.8) {\probs{0}{1}}; 
\node at ( 0   , 2.8) {\probs{\frac{1}{2}}{\frac{1}{2}}}; 
\node at ( 2   , 2.8) {\probs{1}{0}}; 
\node at (-1   , 3.8) {\probs{\frac{1}{5}}{\frac{4}{5}}}; 
\node at ( 1   , 3.8) {\probs{\frac{4}{5}}{\frac{1}{5}}}; 
\node at (-2   , 4.8) {\probs{0}{1}}; 
\node at ( 0   , 4.8) {\probs{\frac{1}{2}}{\frac{1}{2}}}; 
\node at ( 2   , 4.8) {\probs{1}{0}}; 
\node at (-1   , 5.8) {\probs{\frac{3}{14}}{\frac{11}{14}}}; 
\node at ( 1   , 5.8) {\probs{\frac{11}{14}}{\frac{3}{14}}}; 
\node at ( 0   , 6.8) {\probs{\frac{1}{2}}{\frac{1}{2}}}; 

\node at (-1   , 0.2) {\state{0}{1}};
\node at ( 1   , 0.2) {\state{1}{0}};
\node at ( 0  ,  1.2) {\state{1}{1}};
\node at (-2  ,  1.2) {\state{0}{2}};
\node at (-1   , 2.2) {\state{1}{2}};
\node at ( 0.0 , 3.2) {\state{2}{2}};
\node at ( 1   , 2.2) {\state{2}{1}};
\node at ( 2   , 1.2) {\state{2}{0}};
\node at (-2   , 3.2) {\state{1}{3}};
\node at ( 2   , 3.2) {\state{3}{1}};
\node at (-1   , 4.2) {\state{2}{3}};
\node at ( 1   , 4.2) {\state{3}{2}};
\node at ( 0.0 , 5.2) {\state{3}{3}};
\node at (-2   , 5.2) {\state{2}{4}};
\node at ( 2   , 5.2) {\state{4}{2}};
\node at (-1   , 6.2) {\state{3}{4}};
\node at ( 1   , 6.2) {\state{4}{3}};
\node at ( 0.0 , 7.2) {\state{4}{4}};

\node[text=teal] at (-0.5 , 0.7) {$\frac{1}{2}$};
\node[text=teal] at ( 0.5 , 0.7) {$\frac{1}{2}$};
\node[text=teal] at (-1.5 , 1.7) {$\frac{2}{3}$};
\node[text=teal] at (-0.5 , 1.7) {$\frac{1}{3}$};
\node[text=teal] at ( 0.5 , 1.7) {$\frac{1}{3}$};
\node[text=teal] at ( 1.5 , 1.7) {$\frac{2}{3}$};
\node[text=teal] at (-0.5 , 2.7) {$\frac{1}{2}$};
\node[text=teal] at ( 0.5 , 2.7) {$\frac{1}{2}$};
\node[text=teal] at (-1.5 , 3.7) {$\frac{3}{5}$};
\node[text=teal] at (-0.5 , 3.7) {$\frac{2}{5}$};
\node[text=teal] at ( 0.5 , 3.7) {$\frac{2}{5}$};
\node[text=teal] at ( 1.5 , 3.7) {$\frac{3}{5}$};
\node[text=teal] at (-0.5 , 4.7) {$\frac{1}{2}$};
\node[text=teal] at ( 0.5 , 4.7) {$\frac{1}{2}$};
\node[text=teal] at (-1.5 , 5.7) {$\frac{4}{7}$};
\node[text=teal] at (-0.5 , 5.7) {$\frac{3}{7}$};
\node[text=teal] at ( 0.5 , 5.7) {$\frac{3}{7}$};
\node[text=teal] at ( 1.5 , 5.7) {$\frac{4}{7}$};
\node[text=teal] at (-0.5 , 6.7) {$\frac{1}{2}$};
\node[text=teal] at ( 0.5 , 6.7) {$\frac{1}{2}$};

\draw[line width=0.5mm] ( 1,2) -- ( 0,1) node[sloped, pos=0.5, allow upside down]{\arrowIn};
\draw[line width=0.5mm] ( 1,2) -- ( 2,1) node[sloped, pos=0.5, allow upside down]{\arrowIn};
\draw[line width=0.5mm] (-1,2) -- ( 0,1) node[sloped, pos=0.5, allow upside down]{\arrowIn};
\draw[line width=0.5mm] (-1,2) -- (-2,1) node[sloped, pos=0.5, allow upside down]{\arrowIn};
\draw[line width=0.5mm] ( 0,3) -- ( 1,2) node[sloped, pos=0.5, allow upside down]{\arrowIn};
\draw[line width=0.5mm] ( 0,3) -- (-1,2) node[sloped, pos=0.5, allow upside down]{\arrowIn};
\draw[line width=0.5mm] ( 0,1) -- ( 1,0) node[sloped, pos=0.5, allow upside down]{\arrowIn};
\draw[line width=0.5mm] ( 0,1) -- (-1,0) node[sloped, pos=0.5, allow upside down]{\arrowIn};
\draw[line width=0.5mm] ( 1,4) -- ( 0,3) node[sloped, pos=0.5, allow upside down]{\arrowIn};
\draw[line width=0.5mm] ( 1,4) -- ( 2,3) node[sloped, pos=0.5, allow upside down]{\arrowIn};
\draw[line width=0.5mm] (-1,4) -- ( 0,3) node[sloped, pos=0.5, allow upside down]{\arrowIn};
\draw[line width=0.5mm] (-1,4) -- (-2,3) node[sloped, pos=0.5, allow upside down]{\arrowIn};
\draw[line width=0.5mm] ( 0,5) -- ( 1,4) node[sloped, pos=0.5, allow upside down]{\arrowIn};
\draw[line width=0.5mm] ( 0,5) -- (-1,4) node[sloped, pos=0.5, allow upside down]{\arrowIn};
\draw[line width=0.5mm] ( 1,6) -- ( 0,5) node[sloped, pos=0.5, allow upside down]{\arrowIn};
\draw[line width=0.5mm] ( 1,6) -- ( 2,5) node[sloped, pos=0.5, allow upside down]{\arrowIn};
\draw[line width=0.5mm] (-1,6) -- ( 0,5) node[sloped, pos=0.5, allow upside down]{\arrowIn};
\draw[line width=0.5mm] (-1,6) -- (-2,5) node[sloped, pos=0.5, allow upside down]{\arrowIn};
\draw[line width=0.5mm] ( 0,7) -- ( 1,6) node[sloped, pos=0.5, allow upside down]{\arrowIn};
\draw[line width=0.5mm] ( 0,7) -- (-1,6) node[sloped, pos=0.5, allow upside down]{\arrowIn};

\fill ( 1,0) circle[radius=6pt];
\fill (-1,0) circle[radius=6pt];
\fill (-2,1) circle[radius=6pt];
\fill ( 0,1) circle[radius=3pt];
\fill ( 2,1) circle[radius=6pt];
\fill (-1,2) circle[radius=3pt];
\fill ( 1,2) circle[radius=3pt];
\fill (-2,3) circle[radius=6pt];
\fill ( 0,3) circle[radius=3pt];
\fill ( 2,3) circle[radius=6pt];
\fill (-1,4) circle[radius=3pt];
\fill ( 1,4) circle[radius=3pt];
\fill (-2,5) circle[radius=6pt];
\fill ( 0,5) circle[radius=3pt];
\fill ( 2,5) circle[radius=6pt];
\fill (-1,6) circle[radius=3pt];
\fill ( 1,6) circle[radius=3pt];
\fill ( 0,7) circle[radius=3pt];
\end{tikzpicture}

%% file: nnp1_rec.tex
\begin{tikzpicture}[line cap=round,line join=round,x=3cm,y=3cm]
\newcommand{\state}[2]{$(\textcolor{red}{#1},\textcolor{blue}{#2})$}
\newcommand{\probs}[2]{$\left\langle\textcolor{red}{#1};\textcolor{blue}{#2}\right\rangle$}
\newcommand{\half}{\frac{1}{2}}

\draw[line width=0.5mm] ( 1,2) -- ( 0,1) node[sloped, pos=0.5, allow upside down]{\arrowIn};
\draw[line width=0.5mm] ( 1,2) -- ( 2,1) node[sloped, pos=0.5, allow upside down]{\arrowIn};
\draw[line width=0.5mm] (-1,2) -- ( 0,1) node[sloped, pos=0.5, allow upside down]{\arrowIn};
\draw[line width=0.5mm] (-1,2) -- (-2,1) node[sloped, pos=0.5, allow upside down]{\arrowIn};
\draw[line width=0.5mm] ( 0,3) -- ( 1,2) node[sloped, pos=0.5, allow upside down]{\arrowIn};
\draw[line width=0.5mm] ( 0,3) -- (-1,2) node[sloped, pos=0.5, allow upside down]{\arrowIn};

\draw[line width=0.1mm] ( 0,1) -- ( 0,1.3) ;
\draw[line width=0.1mm] ( 1,2) -- ( 1,1.6) ;
\draw[line width=0.1mm] (-1,2) -- (-1,1.6) ;

\fill (-2,1) circle[radius=6pt];
\fill ( 2,1) circle[radius=6pt];
\fill (-1,2) circle[radius=3pt];
\fill ( 1,2) circle[radius=3pt];
\fill ( 0,3) circle[radius=3pt];
\fill ( 0,1) circle[radius=3pt];
\node at ( 2   , 0.8) {\probs{1}{0}};
\node at (-2   , 0.8) {\probs{0}{1}};
\node at ( 0   , 0.8) {\probs{\frac{1}{2}}{\frac{1}{2}}};
\node at ( 1   , 1.5) {\probs{\frac{3n-1}{4n-2}}{\frac{ n-1}{4n-2}}};
\node at (-1   , 1.5) {\probs{\frac{ n-1}{4n-2}}{\frac{3n-1}{4n-2}}};
\node at ( 0   , 2.8) {\probs{\frac{1}{2}}{\frac{1}{2}}};

\node at (-2   , 1.3) {\state{n-2}{n}};
\node at ( 2   , 1.3) {\state{n}{n-2}};
\node at (-1   , 2.3) {\state{n-1}{n}};
\node at ( 1   , 2.3) {\state{n}{n-1}};
\node at ( 0.0 , 3.1) {\state{n}{n}};
\node at ( 0.0 , 1.4) {\state{n-1}{n-1}};

\node[text=teal] at (-0.5 , 1.7) {$\frac{n-1}{2n-1}$};
\node[text=teal] at ( 0.5 , 1.7) {$\frac{n-1}{2n-1}$};
\node[text=teal] at (-1.5 , 1.7) {$\frac{n}{2n-1}$};
\node[text=teal] at ( 1.5 , 1.7) {$\frac{n}{2n-1}$};
\node[text=teal] at (-0.5 , 2.7) {$\half$};
\node[text=teal] at ( 0.5 , 2.7) {$\half$};

\end{tikzpicture}

%% file: nnp2.tex
\begin{tikzpicture}[line cap=round,line join=round,x=2.5cm,y=2.5cm]
\newcommand{\half}{$\frac{1}{2}$}

\newcommand{\state}[2]{$(\textcolor{red}{#1},\textcolor{blue}{#2})$}
\newcommand{\probs}[2]{$\left\langle\textcolor{red}{#1};\textcolor{blue}{#2}\right\rangle$}

\draw[line width=0.5mm] ( 0,1) -- ( 1,0) node[sloped, pos=0.5, allow upside down]{\arrowIn}; 
\draw[line width=0.5mm] ( 0,1) -- (-1,0) node[sloped, pos=0.5, allow upside down]{\arrowIn}; 
\draw[line width=0.5mm] ( 1,2) -- ( 0,1) node[sloped, pos=0.5, allow upside down]{\arrowIn}; 
\draw[line width=0.5mm] ( 1,2) -- ( 2,1) node[sloped, pos=0.5, allow upside down]{\arrowIn}; 
\draw[line width=0.5mm] (-1,2) -- ( 0,1) node[sloped, pos=0.5, allow upside down]{\arrowIn}; 
\draw[line width=0.5mm] (-1,2) -- (-2,1) node[sloped, pos=0.5, allow upside down]{\arrowIn}; 
\draw[line width=0.5mm] ( 0,3) -- ( 1,2) node[sloped, pos=0.5, allow upside down]{\arrowIn}; 
\draw[line width=0.5mm] ( 0,3) -- (-1,2) node[sloped, pos=0.5, allow upside down]{\arrowIn}; 
\draw[line width=0.5mm] ( 1,4) -- ( 0,3) node[sloped, pos=0.5, allow upside down]{\arrowIn}; 
\draw[line width=0.5mm] ( 1,4) -- ( 2,3) node[sloped, pos=0.5, allow upside down]{\arrowIn}; 
\draw[line width=0.5mm] (-1,4) -- ( 0,3) node[sloped, pos=0.5, allow upside down]{\arrowIn}; 
\draw[line width=0.5mm] (-1,4) -- (-2,3) node[sloped, pos=0.5, allow upside down]{\arrowIn}; 
\draw[line width=0.5mm] ( 0,5) -- ( 1,4) node[sloped, pos=0.5, allow upside down]{\arrowIn}; 
\draw[line width=0.5mm] ( 0,5) -- (-1,4) node[sloped, pos=0.5, allow upside down]{\arrowIn}; 
\draw[line width=0.5mm] ( 2,5) -- ( 3,4) node[sloped, pos=0.5, allow upside down]{\arrowIn}; 
\draw[line width=0.5mm] (-2,5) -- (-3,4) node[sloped, pos=0.5, allow upside down]{\arrowIn}; 
\draw[line width=0.5mm] ( 2,5) -- ( 1,4) node[sloped, pos=0.5, allow upside down]{\arrowIn}; 
\draw[line width=0.5mm] (-2,5) -- (-1,4) node[sloped, pos=0.5, allow upside down]{\arrowIn}; 
\draw[line width=0.5mm] ( 2,3) -- ( 1,2) node[sloped, pos=0.5, allow upside down]{\arrowIn}; 
\draw[line width=0.5mm] (-2,3) -- (-1,2) node[sloped, pos=0.5, allow upside down]{\arrowIn}; 
\draw[line width=0.5mm] ( 2,3) -- ( 3,2) node[sloped, pos=0.5, allow upside down]{\arrowIn}; 
\draw[line width=0.5mm] (-2,3) -- (-3,2) node[sloped, pos=0.5, allow upside down]{\arrowIn}; 
\draw[line width=0.5mm] ( 1,6) -- ( 0,5) node[sloped, pos=0.5, allow upside down]{\arrowIn};
\draw[line width=0.5mm] ( 1,6) -- ( 2,5) node[sloped, pos=0.5, allow upside down]{\arrowIn};
\draw[line width=0.5mm] (-1,6) -- ( 0,5) node[sloped, pos=0.5, allow upside down]{\arrowIn};
\draw[line width=0.5mm] (-1,6) -- (-2,5) node[sloped, pos=0.5, allow upside down]{\arrowIn};
\draw[line width=0.5mm] ( 0,7) -- ( 1,6) node[sloped, pos=0.5, allow upside down]{\arrowIn};
\draw[line width=0.5mm] ( 0,7) -- (-1,6) node[sloped, pos=0.5, allow upside down]{\arrowIn};

\node at (-2.0 ,  0.8) {\probs{0}{1}};
\node at ( 1.0 , -0.2) {\probs{1}{0}};
\node at (-1   , -0.2) {\probs{0}{1}};
\node at ( 0   ,  0.8) {\probs{0}{1}};
\node at ( 2   ,  0.8) {\probs{1}{0}};
\node at (-3   ,  1.8) {\probs{0}{1}};
\node at (-1   ,  1.7) {\probs{\frac{1}{6}}{\frac{5}{6}}};
\node at ( 1   ,  1.7) {\probs{\frac{5}{6}}{\frac{1}{6}}};
\node at ( 3   ,  1.8) {\probs{1}{0}};
\node at (-2.0 ,  2.7) {\probs{\frac{1}{24}}{\frac{23}{24}}};
\node at ( 2.0 ,  2.7) {\probs{\frac{23}{24}}{\frac{1}{24}}};
\node at ( 0   ,  2.7) {\probs{\frac{1}{2}}{\frac{1}{2}}};
\node at (-3   ,  3.8) {\probs{0}{1}};
\node at (-1   ,  3.7) {\probs{\frac{9}{40}}{\frac{31}{40}}};
\node at ( 1   ,  3.7) {\probs{\frac{31}{40}}{\frac{9}{40}}};
\node at ( 3   ,  3.8) {\probs{1}{0}};
\node at (-2   ,  4.7) {\probs{\frac{3}{40}}{\frac{37}{40}}};
\node at ( 0   ,  4.7) {\probs{\frac{1}{2}}{\frac{1}{2}}};
\node at ( 2   ,  4.7) {\probs{\frac{37}{40}}{\frac{3}{40}}};
\node at (-1   ,  5.7) {\probs{\frac{9}{35}}{\frac{26}{35}}};
\node at ( 1   ,  5.7) {\probs{\frac{26}{35}}{\frac{9}{35}}};
\node at (-2   ,  6.6) {\probs{\frac{27}{280}}{\frac{253}{280}}};
\node at ( 0   ,  6.7) {\probs{\frac{1}{2}}{\frac{1}{2}}};
\node at ( 2   ,  6.6) {\probs{\frac{253}{280}}{\frac{27}{280}}};
\node at (-1   ,  7.6) {\probs{\frac{139}{504}}{\frac{365}{504}}};
\node at ( 1   ,  7.6) {\probs{\frac{365}{504}}{\frac{139}{504}}};
\node at (-3   ,  5.8) {\probs{0}{1}};
\node at ( 0   ,  8.7) {\probs{\frac{1}{2}}{\frac{1}{2}}};
\node at ( 3   ,  5.8) {\probs{1}{0}};

\node at (-1, 0.2) {\state{0}{1}};
\node at ( 1, 0.2) {\state{1}{0}};
\node at (-2, 1.2) {\state{0}{2}};
\node at ( 0, 1.2) {\state{1}{1}};
\node at ( 2, 1.2) {\state{2}{0}};
\node at (-3, 2.2) {\state{0}{3}};
\node at (-1, 2.2) {\state{1}{2}};
\node at ( 1, 2.2) {\state{2}{1}};
\node at ( 3, 2.2) {\state{3}{0}};
\node at (-2, 3.2) {\state{1}{3}};
\node at ( 0, 3.2) {\state{2}{2}};
\node at ( 2, 3.2) {\state{3}{1}};
\node at (-1, 4.2) {\state{2}{3}};
\node at ( 1, 4.2) {\state{3}{2}};
\node at (-3, 4.2) {\state{1}{4}};
\node at ( 0, 5.2) {\state{3}{3}};
\node at ( 3, 4.2) {\state{4}{1}};
\node at (-2, 5.2) {\state{2}{4}};
\node at ( 2, 5.2) {\state{4}{2}};
\node at (-1, 6.2) {\state{3}{4}};
\node at ( 1, 6.2) {\state{4}{3}};
\node at ( 0, 7.2) {\state{4}{4}};
\node at (-2, 7.2) {\state{3}{5}};
\node at ( 2, 7.2) {\state{5}{3}};
\node at (-1, 8.2) {\state{4}{5}};
\node at ( 1, 8.2) {\state{5}{4}};
\node at ( 0, 9.2) {\state{5}{5}};
\node at ( 3, 6.2) {\state{5}{2}};
\node at (-3, 6.2) {\state{2}{5}};

\node[text=teal] at (-0.5 , 0.7) {\half};
\node[text=teal] at ( 0.5 , 0.7) {\half};
\node[text=teal] at (-1.5 , 1.7) {$\frac{2}{3}$};
\node[text=teal] at (-0.5 , 1.7) {$\frac{1}{3}$};
\node[text=teal] at ( 0.5 , 1.7) {$\frac{1}{3}$};
\node[text=teal] at ( 1.5 , 1.7) {$\frac{2}{3}$};
\node[text=teal] at ( 2.5 , 2.7) {$\frac{3}{4}$};
\node[text=teal] at (-1.5 , 2.7) {$\frac{3}{4}$};
\node[text=teal] at (-0.5 , 2.7) {\half};
\node[text=teal] at ( 0.5 , 2.7) {\half};
\node[text=teal] at ( 1.5 , 2.7) {$\frac{1}{4}$};
\node[text=teal] at (-2.5 , 2.7) {$\frac{3}{4}$};
\node[text=teal] at (-1.5 , 3.7) {$\frac{3}{5}$};
\node[text=teal] at (-0.5 , 3.7) {$\frac{2}{5}$};
\node[text=teal] at ( 0.5 , 3.7) {$\frac{2}{5}$};
\node[text=teal] at ( 1.5 , 3.7) {$\frac{3}{5}$};
\node[text=teal] at (-2.5 , 4.7) {$\frac{4}{6}$};
\node[text=teal] at (-1.5 , 4.7) {$\frac{2}{6}$};
\node[text=teal] at (-0.5 , 4.7) {\half};
\node[text=teal] at ( 0.5 , 4.7) {\half};
\node[text=teal] at ( 1.5 , 4.7) {$\frac{2}{6}$};
\node[text=teal] at ( 2.5 , 4.7) {$\frac{4}{6}$};
\node[text=teal] at (-1.5 , 5.7) {$\frac{4}{7}$};
\node[text=teal] at (-0.5 , 5.7) {$\frac{3}{7}$};
\node[text=teal] at ( 0.5 , 5.7) {$\frac{3}{7}$};
\node[text=teal] at ( 1.5 , 5.7) {$\frac{4}{7}$};

\fill ( 1,0) circle[radius=6pt];
\fill (-1,0) circle[radius=6pt];
\fill (-2,1) circle[radius=6pt];
\fill ( 0,1) circle[radius=3pt];
\fill ( 2,1) circle[radius=6pt];
\fill (-3,2) circle[radius=6pt];
\fill (-1,2) circle[radius=3pt];
\fill ( 1,2) circle[radius=3pt];
\fill ( 3,2) circle[radius=6pt];
\fill (-2,3) circle[radius=3pt];
\fill ( 0,3) circle[radius=3pt];
\fill ( 2,3) circle[radius=3pt];
\fill (-3,4) circle[radius=6pt];
\fill (-1,4) circle[radius=3pt];
\fill ( 1,4) circle[radius=3pt];
\fill ( 3,4) circle[radius=6pt];
\fill (-2,5) circle[radius=3pt];
\fill ( 0,5) circle[radius=3pt];
\fill ( 2,5) circle[radius=3pt];
\fill (-1,6) circle[radius=3pt];
\fill ( 1,6) circle[radius=3pt];
\fill ( 0,7) circle[radius=3pt];


\draw[line width=0.5mm] (-2,7) -- (-3,6) node[sloped, pos=0.5, allow upside down]{\arrowIn};
\draw[line width=0.5mm] ( 2,7) -- ( 3,6) node[sloped, pos=0.5, allow upside down]{\arrowIn};
\draw[line width=0.5mm] (-2,7) -- (-1,6) node[sloped, pos=0.5, allow upside down]{\arrowIn};
\draw[line width=0.5mm] ( 2,7) -- ( 1,6) node[sloped, pos=0.5, allow upside down]{\arrowIn};
\draw[line width=0.5mm] ( 1,8) -- ( 0,7) node[sloped, pos=0.5, allow upside down]{\arrowIn};
\draw[line width=0.5mm] ( 1,8) -- ( 2,7) node[sloped, pos=0.5, allow upside down]{\arrowIn};
\draw[line width=0.5mm] (-1,8) -- ( 0,7) node[sloped, pos=0.5, allow upside down]{\arrowIn};
\draw[line width=0.5mm] (-1,8) -- (-2,7) node[sloped, pos=0.5, allow upside down]{\arrowIn};
\draw[line width=0.5mm] ( 0,9) -- ( 1,8) node[sloped, pos=0.5, allow upside down]{\arrowIn};
\draw[line width=0.5mm] ( 0,9) -- (-1,8) node[sloped, pos=0.5, allow upside down]{\arrowIn};

\node[text=teal] at (-2.5 , 6.7) {$\frac{5}{8}$};
\node[text=teal] at (-1.5 , 6.7) {$\frac{3}{8}$};
\node[text=teal] at (-0.5 , 6.7) {\half};
\node[text=teal] at ( 0.5 , 6.7) {\half};
\node[text=teal] at ( 1.5 , 6.7) {$\frac{3}{8}$};
\node[text=teal] at ( 2.5 , 6.7) {$\frac{5}{8}$};
\node[text=teal] at (-1.5 , 7.7) {$\frac{5}{9}$};
\node[text=teal] at (-0.5 , 7.7) {$\frac{4}{9}$};
\node[text=teal] at ( 0.5 , 7.7) {$\frac{4}{9}$};
\node[text=teal] at ( 1.5 , 7.7) {$\frac{5}{9}$};
\node[text=teal] at (-0.5 , 8.7) {\half};
\node[text=teal] at ( 0.5 , 8.7) {\half};

\fill ( 3,6) circle[radius=6pt];
\fill (-3,6) circle[radius=6pt];
\fill ( 2,7) circle[radius=3pt];
\fill (-2,7) circle[radius=3pt];
\fill ( 1,8) circle[radius=3pt];
\fill (-1,8) circle[radius=3pt];
\fill ( 0,9) circle[radius=3pt];

\end{tikzpicture}

%% file: nnp2_cell.tex
\begin{tikzpicture}[line cap=round,line join=round,x=3cm,y=3cm]

\newcommand{\half}{$\frac{1}{2}$}
\newcommand{\state}[2]{$(\textcolor{red}{#1},\textcolor{blue}{#2})$}
\newcommand{\probs}[2]{$\left\langle\textcolor{red}{#1};\textcolor{blue}{#2}\right\rangle$}

\draw[line width=0.5mm] ( 1,1) -- ( 0,0) node[sloped, pos=0.5, allow upside down]{\arrowIn}; 
\draw[line width=0.5mm] (-1,1) -- ( 0,0) node[sloped, pos=0.5, allow upside down]{\arrowIn}; 
\draw[line width=0.5mm] ( 0,2) -- ( 1,1) node[sloped, pos=0.5, allow upside down]{\arrowIn}; 
\draw[line width=0.5mm] ( 0,2) -- (-1,1) node[sloped, pos=0.5, allow upside down]{\arrowIn}; 
\draw[line width=0.5mm] ( 1,1) -- ( 2,0) node[sloped, pos=0.5, allow upside down]{\arrowIn};

\draw[line width=0.1mm] (0,1.6) -- (0,2) ;
\draw[line width=0.1mm] (1,0.7) -- (1,1) ;
\draw[line width=0.1mm] (2,0.3) -- (2,0) ;

\fill ( 0,0) circle[radius=3pt];
\fill (-1,1) circle[radius=3pt];
\fill ( 1,1) circle[radius=3pt];
\fill ( 0,2) circle[radius=3pt];
\fill ( 2,0) circle[radius=6pt];

\node at ( 0   , -0.1){\probs{x_{n}}{1-x_{n}}};
\node at (-1   , 0.8) {\probs{\frac{1}{2}}{\frac{1}{2}}};
\node at ( 0   , 1.5) {\probs{x_{n+1}}{1-x_{n+1}}};
\node at ( 1 , 1.3) {\probs{y}{1-y}};
\node at ( 2   , -0.2) {\probs{1}{0}};

\node at ( 0  , 0.3) {\state{n}{n-1}};
\node at ( 2  , 0.4) {\state{n+2}{n-2}};
\node at ( 0  ,  2.1 ) {\state{n+1}{n}};
\node at (-1 ,   1.2 ) {\state{n}{n}};
\node at ( 1 ,   0.6 ) {\state{n+1}{n-1}};

\node[text=teal] at (-0.5 , 1.7) {$\frac{n}{2n+1}$};
\node[text=teal] at ( 0.5 , 1.7) {$\frac{n+1}{2n+1}$};
\node[text=teal] at (-0.5 , 0.7) {$\frac{1}{2}$};
\node[text=teal] at ( 0.5 , 0.7) {$\frac{n-1}{2n}$};
\node[text=teal] at ( 1.5 , 0.7) {$\frac{n+1}{2n}$};

\end{tikzpicture}

%% file: nnp3_cell.tex
\begin{tikzpicture}[line cap=round,line join=round,x=3cm,y=3cm]
\newcommand{\half}{$\frac{1}{2}$}
\newcommand{\state}[2]{$(\textcolor{red}{#1},\textcolor{blue}{#2})$}
\newcommand{\probs}[2]{$\left\langle\textcolor{red}{#1};\textcolor{blue}{#2}\right\rangle$}

\draw[line width=0.5mm] ( 1,1) -- ( 0,0) node[sloped, pos=0.5, allow upside down]{\arrowIn}; 
\draw[line width=0.5mm] (-1,1) -- ( 0,0) node[sloped, pos=0.5, allow upside down]{\arrowIn}; 
\draw[line width=0.5mm] ( 0,2) -- ( 1,1) node[sloped, pos=0.5, allow upside down]{\arrowIn}; 
\draw[line width=0.5mm] ( 0,2) -- (-1,1) node[sloped, pos=0.5, allow upside down]{\arrowIn}; 
\draw[line width=0.5mm] ( 1,1) -- ( 2,0) node[sloped, pos=0.5, allow upside down]{\arrowIn}; 
\draw[line width=0.5mm] (-1,1) -- (-2,0) node[sloped, pos=0.5, allow upside down]{\arrowIn}; 

\draw[line width=0.1mm] (0,1.5) -- (0,2);

\fill ( 0,0) circle[radius=3pt];
\fill (-1,1) circle[radius=3pt];
\fill ( 1,1) circle[radius=3pt];
\fill ( 0,2) circle[radius=3pt];
\fill ( 2,0) circle[radius=6pt];
\fill (-2,0) circle[radius=3pt];

\node at ( 0 , -0.2) {\probs{x_{n}}{1-x_{n}}};
\node at (-1 , 0.7) {\probs{y}{1-y}};
\node at ( 0   , 1.4) {\probs{x_{n+1}}{1-x_{n+1}}};
\node at ( 1 , 0.7) {\probs{z}{1-z}};
\node at ( 2 , -0.2) {\probs{1}{0}};
\node at (-2   , -0.2) {\probs{\frac{1}{2}}{\frac{1}{2}}};

\node at ( 0  , 0.3) {\state{n}{n-1}};
\node at ( 2  , 0.4) {\state{n+1}{n-3}};
\node at ( 0  ,  2.1 ) {\state{n+1}{n-1}};
\node at (-1 ,   1.3 ) {\state{n}{n-1}};
\node at (-2 ,   0.4 ) {\state{n-1}{n-1}};
\node at ( 1 ,   1.4 ) {\state{n+1}{n-2}};

\node[text=teal] at (-0.5 , 1.7) {$\frac{n-1}{2n}$};
\node[text=teal] at ( 0.5 , 1.7) {$\frac{n+1}{2n}$};
\node[text=teal] at (-0.5 , 0.7) {$\frac{n}{2n-2}$};
\node[text=teal] at ( 0.5 , 0.7) {$\frac{n-2}{2n-1}$};
\node[text=teal] at ( 1.5 , 0.7) {$\frac{n+1}{2n-1}$};
\node[text=teal] at (-1.5 , 0.7) {$\frac{n-1}{2n-2}$};

\end{tikzpicture}